\documentclass[aps,twocolumn,pra,superscriptaddress,amsmath,amssymb]{revtex4-2}
\usepackage{dcolumn}
\usepackage{bm}
\usepackage{amsmath}
\usepackage{txfonts}
\usepackage[T1]{fontenc}
\usepackage{xspace}
\usepackage{ulem}
\usepackage{comment}
\usepackage{braket}
\ifx\pdfoutput\undefined
\usepackage[dvipdfmx]{graphicx}
\usepackage[dvipdfmx]{hyperref}
\usepackage[dvipdfmx]{color}
\else
\usepackage{graphicx}
\usepackage{hyperref}
\usepackage{color}
\fi

\begin{document}
\let\emph\textit

\title{
  Critical behavior of the Ising model on square-triangle tilings
}

\author{Akihisa Koga}
\affiliation{
  Department of Physics, Institute of Science Tokyo,
  Meguro, Tokyo 152-8551, Japan
}

\author{Shiro Sakai}
\affiliation{
  Center for Emergent Matter Science, RIKEN, Wako, Saitama 351-0198, Japan
}

\date{\today}
\begin{abstract}
  We investigate magnetic properties of the ferromagnetic Ising model
  on square-triangle tilings to explore
  how the hyperuniformity, which characterizes long-range behavior of the point pattern,
  influences critical phenomena where long-range correlations play a crucial role.
  The square-triangle tilings are spatially random structures in two dimensions
  constructed
  by densely packing the plane with squares and triangles.
  The growth rule with a parameter $p$ proposed in our previous paper
  enables systematic generations of 
  hyperuniform, nonhyperuniform, and antihyperuniform tilings.
  Classical Monte Carlo simulations of the Ising model on these tilings 
  show that critical behavior always belongs to
  the two-dimensional Ising universality class.
  It is clarified that the critical temperature
  is higher for the tiling with higher regularity in terms of hyperuniformity.
  Critical phenomena in the Ising models
  on the periodic and quasiperiodic tilings composed of the square and triangle tiles
  are also addressed.
\end{abstract}

\maketitle

\section{Introduction}

Hyperuniformity has attracted considerable interest as
a framework for classifying point distributions based on their long-range behavior~\cite{Torquato_2003,Torquato_2018}.
Specially, point patterns can be categorized as hyperuniform, nonhyperuniform, and anti-hyperuniform,
depending on density fluctuations of the point pattern in a large scale.
They are measured by the variance of the number of points within 
a spherical observation window of radius $R$.
When we write the variance in the form of $\sigma^2(R)\propto R^{d-\kappa}$,
where $d$ is the spatial dimension and $\kappa$ is constant,
point patterns are classified as hyperuniform for $\kappa>0$,
non-hyperuniform for $\kappa=0$, and antihyperuniform for $\kappa<0$.
Hyperuniform point patterns can be further divided into distinct classes: Class I (III) for $\kappa=1\; (0<\kappa<1)$, 
and Class II for the variance scaling as $\sigma^2(R)\propto R^{d-1}\log R$.
Periodic and perfect quasiperiodic point patterns belong to Class I.
In contrast, the random point patterns are nonhyperuniform, 
while fractal point patterns are anti-hyperuniform.
These classifications arise from the long-range behavior
inherent in each point distribution.
It is interesting to consider correlated electron or spin systems
on such distinct structures.
In fact, recent studies~\cite{Fuchs_2019,Sakai_2022,Sakai_2022_2,Hori_2024}
have found hyperuniform distributions of local physical quantities
in quasiperiodic lattices.
Then, a natural question arises:
do hyperuniform properties in the point distributions affect
critical phenomena in the strongly correlated systems 
where long range correlations play an essential role?

One of the simplest toy models for discussing cooperative phenomena
is the two-dimensional ferromagnetic Ising model.
The Ising model on certain periodic lattices
is exactly solvable~\cite{Onsager_1944,Wannier_1945},
and it is known that the critical temperature to
the ferromagnetically ordered state 
depends on the geometry of the lattice~\cite{Thompson_Wardrop_1974,Codello_2010}.
However, critical behavior remains within the same two-dimensional Ising universality class
across different lattice geometries.
In addition to the periodic lattices, the Ising model has also been numerically examined on quasiperiodic lattices
such as Penrose~\cite{Bhattacharjee_Ho_Johnson_1987,Okabe_Niizeki_1988,So_1991,Ma_2004,Komura_2016,Azhari}
and octagonal tilings~\cite{Ledue_Teillet_1995,Ledue_1996},
as well as on randomly distributed lattices~\cite{Janke,Lima}.
Despite differences in the critical temperatures, 
it has been clarified that the critical phenomena in these systems belong
to the same two-dimensional Ising universality class.
While the critical temperature generally depends on 
the number of exchange couplings between the nearest neighbor sites, it is not solely determined by it.
This suggests that the lattice structure, in particular, its hyperuniform property,
plays a crucial role in influencing the critical phenomena.
Therefore, systematic studies are necessary to investigate critical phenomena 
of the Ising model with distinct hyperuniform properties.

One of the potential candidates for a platform of such a study is 
the square-triangle tiling~\cite{Collins_1964,Kawamura_1983,HenleyBook,Oxborrow_Henley_1993,Xiao_2012,Clerc_2021},
which is a spatially random structure in two dimensions constructed by densely 
packing the plane with regular square and triangle tiles.
In our previous paper~\cite{Koga}, 
we have proposed a growth rule with a parameter $p$
to generate the square-triangle tilings
with distinct hyperuniform properties.
The tilings with $0<p<p_c$ are classified as anti-hyperuniform with $\kappa<0$, 
while those with $p_c<p<1$ are hyperuniform with $\kappa>0$, where $p_c\sim 0.5$.
Furthermore, periodic and quasiperiodic tilings exist within this family,
which belong to the Class I with $\kappa=1$.
Thus, this tiling comprehensively includes elements such as hyperuniform, nonhyperuniform, and antihyperuniform structures. 
The Ising model on the square-triangle tiling provides an appropriate platform
to elucidate the relationship between critical phenomena and hyperuniform properties.

In this paper, we consider the square-triangle tilings with distinct hyperuniform properties
and study magnetic properties of the Ising model on them.
We apply the classical Monte Carlo (MC) simulations to the ferromagnetic Ising model on the square-triangle tilings
of various system sizes and determine the critical temperatures.
We then demonstrate that the magnetization and susceptibility
are properly scaled with the exponents of the two-dimensional Ising universality class.
We address how the critical temperature is related to the exponent $\kappa$
and order metric in the framework of the hyperuniformity.

This paper is organized as follows.
In Sec.~\ref{model},
we introduce the two-dimensional Ising model
on the square-triangle tilings,
which are generated by the growth rule.
In Sec.~\ref{results},
we discuss critical behavior of the Ising model,
examining the scaling functions
and elucidate the correlation between the critical temperature
and regularity of the structure.
A summary is given in the last section.

\section{Model and method}\label{model}

\begin{figure}[htb]
  \includegraphics[width=\linewidth]{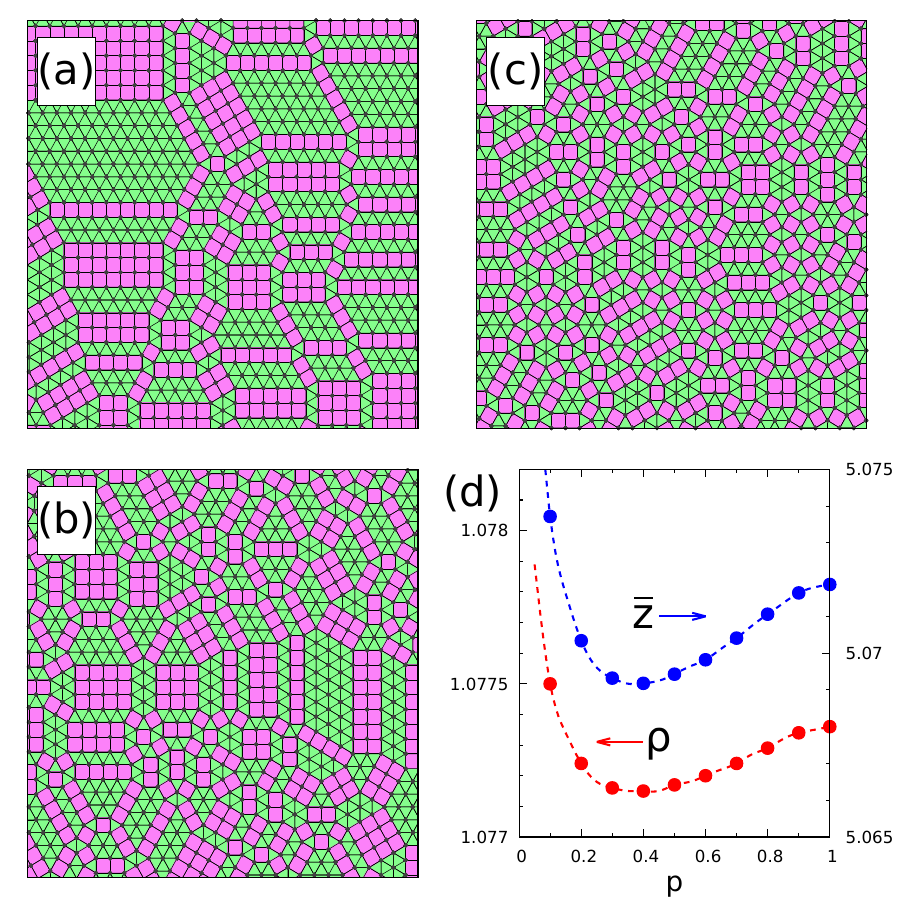}
  \caption{
    Square-triangle tilings generated by the growth rule
    with (a) $p=0.2$, (b) $p=0.5$, and (c) $p=1.0$.
    (d) Point density $\rho$ and average coordination number $\bar{z}$
    as a function of $p$.
  }
  \label{lattice}
\end{figure}

We consider the Ising model on the square-triangle tiling
composed of the squares and triangles~\cite{Collins_1964,Kawamura_1983}.
In this tiling, the local configurations are limited to
four distinct types of vertices,
which are referred to as $3^6$, $3^2434$, $3^34^2$, and $4^4$,
according to the numbers (denoted by superscripts) of triangles (3) and squares (4) sharing a vertex.
In this work, we generate the tilings
by means of a growth rule~\cite{Koga},
where square and triangle tiles are iteratively attached
on the surface of the finite tiling.
As elaborated in Ref.~\cite{Koga}, we introduce a probability $p$ $(0\le p\le 1)$ 
when a pair of square and triangle tiles is attached to a $3^24$ vertex on the surface: 
a $3^2434$ vertex is generated at the probability $p$ 
while a $3^34^2$ vertex is generated at the probability $1-p$. 
This simple rule can control hyperuniform properties of the square-triangle tilings~\cite{Koga}.
Figure~\ref{lattice} shows the square-triangle tilings generated 
by the growth rule with $p=0.2, 0.5$, and $1.0$.
When $p$ is small, large square and triangular domains,
which are composed of adjacent square and triangle tiles,
are distributed, and $3^6$ and $4^4$ vertices frequently appear 
in the tiling, as shown in Fig.~\ref{lattice}(a).
In this case, the variance of the point pattern shows $\kappa<0$ in the long range region and 
this point pattern is classified as antihyperuniform.
On the other hand, when $p$ is large,
larger domains diminish, resulting in a more uniform mixture of squares and triangles,
with the $3^2434$ and $3^34^2$ vertices being dominant in the tiling,
as shown in Fig.~\ref{lattice}(c).
In the case with $p_c<p\; (p_c\sim 0.5)$,
the point patterns are hyperuniform with $0<\kappa<1$.
The tiling structure at the transition point $(p=p_c)$
is shown in Fig.~\ref{lattice}(b).
Figure~\ref{lattice}(d) shows the point density $\rho$ and average coordination number $\bar{z}$
as a function of $p$.
We find that these quantities change little for $0.2\le p \le 1.0$.
On the other hand, when $p\le 0.2$, both quantities take large values.
This may mean that large triangular domains become dominant.
Therefore, the phase separation is expected in these tilings, 
leading to complex magnetic properties
when $p\rightarrow 0$.
In the following, we focus on the cases with $0.2\le p\le 1.0$.

We treat the Ising model on the square-triangle tiling given 
by the following Hamiltonian
\begin{align}
  H=-J\sum_{(ij)}S_iS_j,
\end{align}
where $S_i(=\pm 1)$ is the classical spin at the $i$th site and
$J(>0)$ is the coupling constant.
The summation is taken over nearest-neighbor site pairs $(ij)$, 
which are connected by the edges of tiles in the square-triangle tilings.
The Ising models on two-dimensional periodic and quasiperiodic tilings
have been examined analytically and numerically~\cite{Onsager_1944,Wannier_1945,Thompson_Wardrop_1974,Codello_2010,Bhattacharjee_Ho_Johnson_1987,Okabe_Niizeki_1988,So_1991,Ma_2004,Komura_2016,Azhari,Ledue_Teillet_1995,Ledue_1996}.
It has been clarified that
the critical phenomena belong to the two-dimensional Ising universality class,
while the transition temperature depends on the geometry of the lattice.

The classical MC method is appropriate to study
thermodynamic properties in the Ising model on general lattices.
It is known that the square-triangle tiling is topologically equivalent to
the square lattice with dilute diagonal bonds~\cite{Kawamura_1983}.
Therefore, the lattice sites in the square-triangle tiling can be
divided into three sublattices.
This enables us to use
the sublattice update algorithm in the simulations.
Nevertheless, the global update algorithm is more useful than
the sublattice one in the critical region.
In our calculations,
we use Swendsen-Wang (SW) multi-cluster algorithm~\cite{SW} for the global update.
We have confirmed that the numerical results obtained by means of
these two algorithms agree with each other, and 
the SW algorithm is more efficient than the other
around the critical temperature.

We perform the MC simulations with SW algorithm 
to calculate
the ensemble average of absolute value of magnetization $m$,
magnetic susceptibility $\chi$, and
the Binder parameter $U$~\cite{Binder},
which are defined as
\begin{align}
M&=\left|\sum_i S_i\right|,\\
m&=\frac{1}{N}\langle M\rangle,\\
\chi&=\frac{1}{NT}\left(\langle M^2\rangle-\langle M\rangle^2\right),\\
U&=1-\frac{\langle M^4\rangle}{3\langle M^2\rangle^2},
\end{align}
where $\langle A \rangle$ is the ensemble average of the quantity $A$,
$M$ is the total magnetization, $N$ is the total number of sites, and $T$ is the temperature.

Here, we comment on the boundary condition of the lattice.
In general, a periodic boundary condition (PBC) is usually adopted in simulating the Ising model. 
However, generating the square-triangle tiling with PBC 
by means of the growth rule presents significant challenges.
Alternatively, the Ising model with open boundary condition (OBC)
is also applicable to study critical phenomena.
In fact, the Ising model on the Penrose tiling 
has successfully been investigated with the OBC~\cite{Azhari}.
In this study, we focus on circular region with a radius $R$ ($N\sim \rho \pi R^2$)
in the square-triangle tiling with OBC.

When examining critical phenomena, it is crucial to consider
the dependence of the physical quantities on the system size.
In this study, we consider the circular regions with distinct radii.
We also address the sample dependence
in the square-triangle tilings, which arise from the growth rule
where square and triangle tiles are randomly attached in the growing process.
For circular regions with the radius $R$,
the number of possible configurations scales as $\exp (cR^2)$
where $c$ is a positive constant.
This may make it difficult to obtain the numerically exact results for a given $R$.
In our calculations, we fix the tiling structure instead of the radius $R$.
We first generate a single large tiling and 
cut out several circular regions with distinct radii, all sharing a common center.
Subsequently, we evaluate physical quantities of the Ising model on each circular region
by means of the MC simulations.
Finally, we apply the finite size scaling technique to the MC results 
and examine critical phenomena for this single large tiling.
By performing similar analyses for several different large tilings
and averaging the results,
we discuss the critical phenomena in the Ising model on the square-triangle tilings.

\section{Results}\label{results}

\begin{figure}[htb]
  \includegraphics[width=0.8\linewidth]{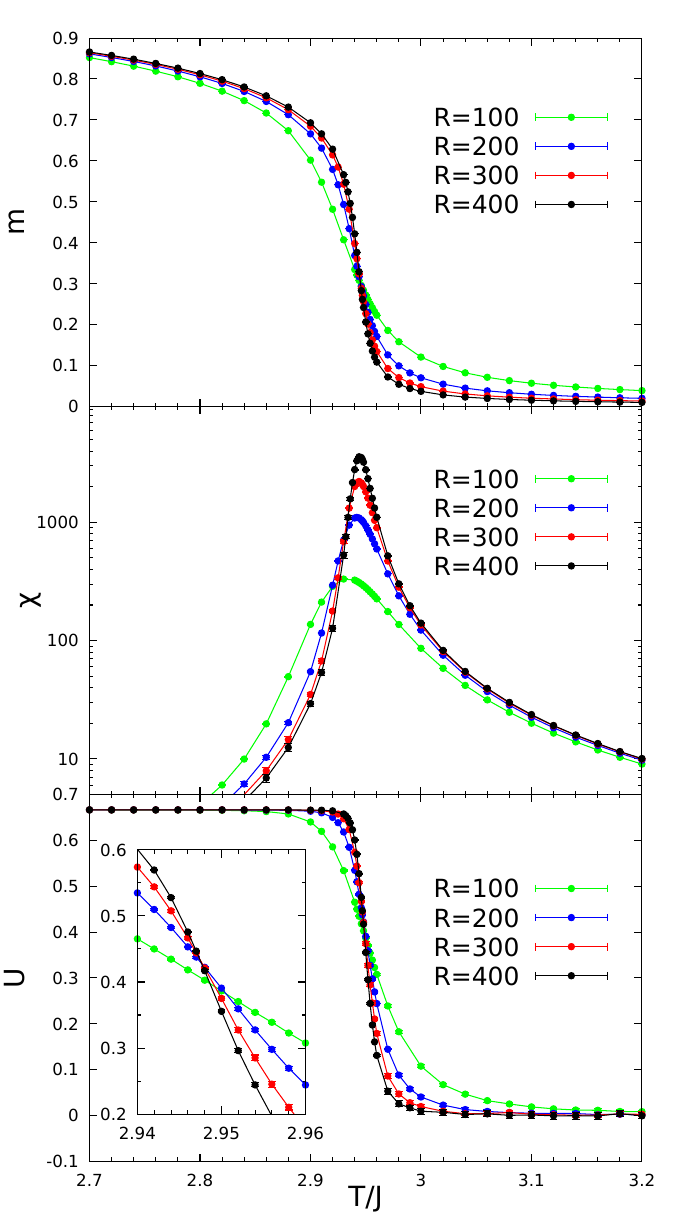}
  \caption{
  Magnetization, susceptibility, and Binder parameter
  as a function of temperature 
  for the ferromagnetic Ising model on the square-triangle tilings with $p=0.5$.
  }
  \label{R0.5}
\end{figure}

We first consider the Ising model on the square-triangle tiling
with $p=0.5$.
It is known that the vertex structure on this tiling
is nonhyperuniform $(\kappa\sim 0)$~\cite{Koga}.
Therefore, this vertex structure is different in long-range behavior 
from any conventional periodic and quasiperiodic lattices,
whose point distributions are hyperuniform $(\kappa=1)$.
Performing the MC simulations for the circular regions
with $R=100, 200, 300$, and $400$,
we obtain the magnetization, magnetic susceptibility,
and Binder parameter, 
which are shown in Fig.~\ref{R0.5}.
At high temperatures, the magnetization is almost zero, and 
the system is paramagnetic.
Decreasing temperatures, around $T/J=2.95$, the magnetization rapidly increases
and the magnetic susceptibility shows a peak.
Furthermore, we find that
the corresponding singularities become stronger 
with increasing $R$.
This suggests the existence of the second-order phase transition
to the ferromagnetically ordered state with the spontaneous magnetization.

In our simulations, we have considered the circular regions with OBC.
Since the structure is not deterministic,
it is generally hard 
to accurately evaluate the physical quantities
in the thermodynamic limit at any temperature.
Nevertheless, the critical temperature can be evaluated quantitatively 
by means of the Binder parameter.
The bottom panel of Fig.~\ref{R0.5} shows that
the lines of the Binder parameter for distinct radii cross at almost the same point.
Therefore, we can evaluate the critical temperature $T_c/J=2.948$ in this case.
The slight difference in the crossing points between two lines originates from 
the sample dependence.
It is expected that, as $R$ increases,
bulk properties less sensitive to the microscopic structure become dominant. 
Consequently, the critical temperature can be determined with greater precision.

\begin{figure*}[htb]
  \includegraphics[width=\linewidth]{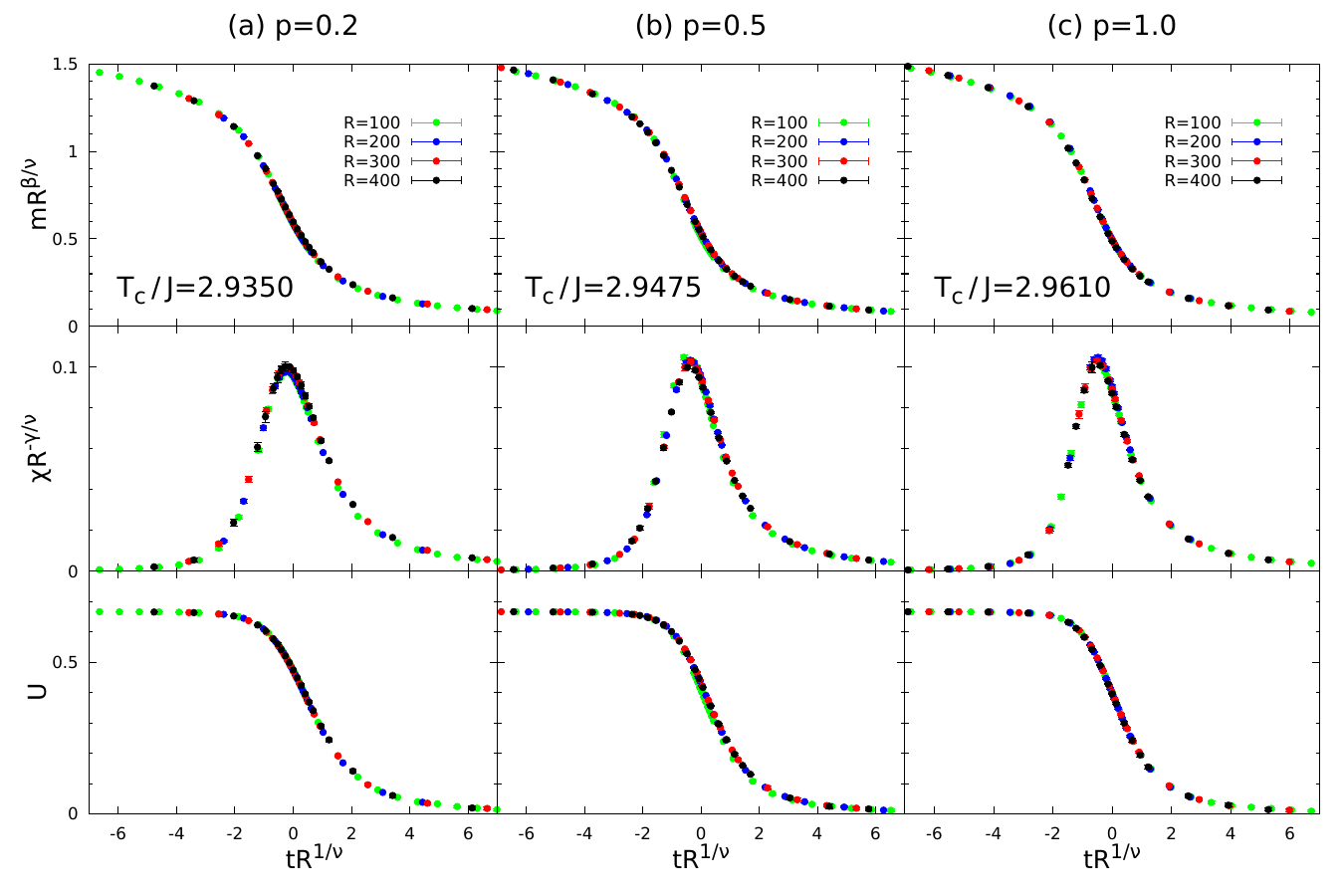}
  \caption{
  Rescaled magnetization $mR^{\beta/\nu}$, susceptibility $\chi R^{-\gamma/\nu}$, and Binder parameter $U$
  as a function of rescaled temperature $tR^{1/\nu}$ with $t=(T-T_c)/T_c$
  for the ferromagnetic Ising model on the square-triangle tilings with (a) $p=0.2$, (b) $p=0.5$, and (c) $p=1.0$.
  Here, we have used the critical exponents of the two-dimensional Ising universality class.
  }
  \label{scaling}
\end{figure*}

Next, we discuss critical behavior in this magnetic phase transition.
It is known that 
the magnetization, susceptibility, and Binder parameter are scaled 
around the critical temperature $T_c$ as
\begin{align}
  m(t,L)&=L^{-\beta/\nu}\tilde{m}(tL^{1/\nu}),\\
  \chi(t,L)&=L^{\gamma/\nu}\tilde{\chi}(tL^{1/\nu}),\\
  U(t,L)&=\tilde{U}(tL^{1/\nu}),
\end{align}
where $\tilde{m}, \tilde{\chi},$ and $\tilde{U}$ are the rescaled functions
for the magnetization, susceptibility, and Binder parameter, respectively.
$\beta, \gamma$, and $\nu$ are the critical exponents, 
$t[=(T-T_c)/T_c]$ is the rescaled temperature and $L$ is the characteristic length 
of the system.
It has been clarified that the phase transition of the Ising models 
on the periodic and quasiperiodic tilings always belongs to
the two-dimensional Ising universality class
with critical exponents $\beta=1/8, \gamma=7/4$, and $\nu=1$.
To clarify whether or not the phase transitions in our model also
belong to this universality class,
we calculate the scaling functions for magnetization, susceptibility,
and Binder parameter by means of the above exponents.
The results with $p=0.2, 0.5$, and $1.0$
are shown in Fig.~\ref{scaling}.
Here, we have used the radius $R$ of the circular region as the length scale $L$ of the system and
the critical temperatures $T_c/J=2.9350, 2.9475,$ and $2.9610$
have been determined from the crossing point of Binder parameters
for the tilings with $p=0.2, 0.5$, and $1.0$, respectively. 
We find in Fig.~\ref{scaling} that these values give fairly good scaling plots.
These mean that the magnetic phase transitions in the Ising model on the square-triangle tilings,
which are hyperuniform, nonhyperuniform, or antihyperuniform,
always belong to the two dimensional Ising universality class.

\begin{figure}[htb]
  \includegraphics[width=\linewidth]{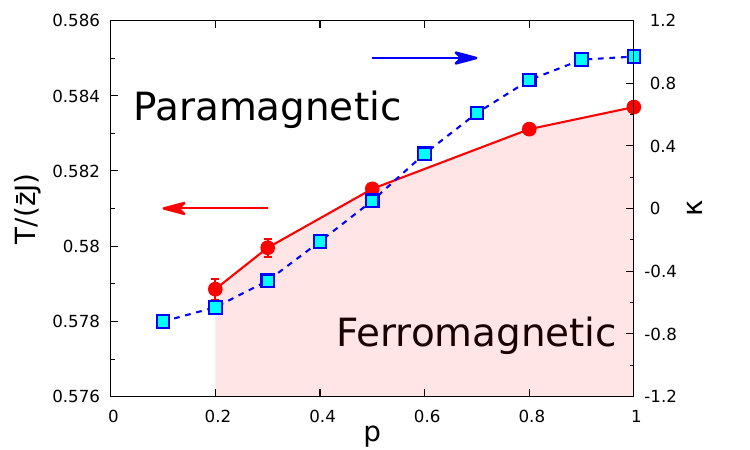}
  \caption{
  Phase diagram of the Ising model on the square-triangle tilings with $p$.
  Red circles denote the normalized critical temperature 
  while blue squares denote the exponent of the point-density variance.
  }
  \label{PD}
\end{figure}

We generate more than five distinct large tilings for each $p$,
perform MC simulations, 
and determine the transition temperature
by deducing the crossing point of the lines for the Binder parameters
with $R=300$ and $400$.
The phase diagram is shown in Fig.~\ref{PD}.
We find that as $p$ increases,
the transition temperature normalized by the average coordination number $T_c/(\bar{z}J)$
slightly increases.
In general, the transition temperature depends on
the coordination number of the lattice
since the energy should be proportional to the number of bonds. 
In our square-triangle tilings,
the coordination number changes only
by less than 0.1 percent for $0.2\le p\le 1.0$, as shown in Fig.~\ref{lattice}(d).
Therefore, in this case, the change in the coordination number has a minor effect on 
the change in the critical temperature
and the major effect is due to the hyperuniform properties
i.e., the exponent $\kappa$ of the variance of the number of points.
As we can see in Fig.~\ref{PD}, $\kappa$ increases with $p$, 
meaning that the structure acquires more regularity. 
Concomitantly with this, the normalized critical temperature increases.

Finally, we compare our results with the critical temperatures 
of the Ising model on the periodic and quasiperiodic lattices
composed of square and triangle tiles.
Magnetic properties of the Ising model on various lattices,
including square, triangular, trellis, and Shastry-Sutherland lattices
have been examined in previous studies~\cite{Onsager_1944,Wannier_1945,Codello_2010}.
For the Stampfli hexagonal and dodecagonal tilings,
we present new results obtained 
by means of the classical MC simulations,
with further details provided in Appendix~\ref{Stampfli}.
These results are summarized in Table~\ref{tbl}.
\begin{table*}[htb]
  \caption{Vertex fractions $f_i \;(i=3^6, 3^2434, 3^34^2, 4^4)$,
    point density $\rho$, ratio of numbers of triangles and squares $N_\triangle/N_\square$,
    average of the coordination number $\bar{z}$,
    critical temperature of the Ising model $T_c/J$,
    and order metric $B$~\cite{Torquato_2003,KogaSakai}
    for various tilings composed of squares and triangles.
    }
  {
  \renewcommand\arraystretch{1.2}
  \begin{tabular}{c|cccc|ccccc|c}
    \toprule
    Lattice & $f_{3^6}$ & $f_{3^2434}$ & $f_{3^34^2}$ & $f_{4^4}$ & $\rho$ & $N_\triangle/N_\square$ & $\bar{z}$ & $T_c/J$ & $T_c/(\bar{z}J)$ & $B$ \\
    \hline
    Triangular & 1 & 0 & 0 & 0 & $1.155$ & $\infty$ & 6 & 3.6410 & 0.6068 & 0.508347\\
    Shastry-Sutherland & 0 & 1 & 0 & 0 & $1.072$ & 2 & 5 & 2.9263 & 0.5853 & 0.51664\\
    Trellis & 0 & 0 & 1 & 0 & $1.072$ & 2 & 5& 2.8854 & 0.5771 & 0.51877\\
    Square & 0 & 0 & 0 & 1 & 1 & 0 & 4 & 2.2692 & 0.5673 & 0.516401 \\
    \hline
    hexagonal Stampfli & 0.071 & 0.746 & 0.182 & 0 & $1.077$ & 2.309 & 5.072 & 2.9692 & 0.5854 & 0.51785\\
    dodecagonal Stampfli & 0.071 & 0.746 & 0.182 & 0 & $1.077$ & 2.309 & 5.072 & 2.9725 & 0.5862 &0.51764 \\
    \hline
    Our tiling ($p=0.2$) & 0.23 & 0.15 & 0.45 & 0.16 & 1.077 & 2.29 & 5.07 & 2.935 & 0.579 \\
    Our tiling ($p=0.5$) & 0.15 & 0.32 & 0.45 & 0.08 & 1.077 & 2.29 & 5.07 & 2.948 & 0.581 \\
    Our tiling ($p=1$)   & 0.08 & 0.57 & 0.34 & 0.01 & 1.077 & 2.31 & 5.07 & 2.961 & 0.584 \\
    \toprule
  \end{tabular}
  }
  \label{tbl}
\end{table*}
To facilitate comparisons across different lattice models,
we first use the normalized critical temperature $T_c/(\bar{z}J)$.
After this normalization, the triangular lattice has the highest critical temperature, 
while the square lattice has the lowest. 
The critical temperatures of other tilings composed of square and triangle tiles fall between these extremes, 
suggesting that vertex density $\rho$ and coordination number $\bar{z}$ 
significantly influence the normalized critical temperature.
However, even for the lattices with the same $\rho$ and $\bar{z}$
such as the Shastry-Sutherland and trellis lattices, 
as well as the Stampfli hexagonal and dodecagonal tilings,
the critical temperatures differ.
This indicates that, in addition to $\rho$ and $\bar{z}$, the ``regularity'' of the tiling structure
plays a secondary but significant role.

Here, we introduce the order metric $B$ defined in Ref.~\cite{Torquato_2003},
characterizing the Class I hyperuniform point pattern,
which is given as
\begin{align}
  B=\phi^{-(d-1)/d}\lim_{R\rightarrow\infty}\frac{\sigma^2(R)}{R^{d-1}},
\end{align}
with $\phi=\pi \rho / 4$.
A systematic study of the point patterns exhibits that
the smaller order metric should
indicate high regularity~\cite{Torquato_2003,Torquato_2018,KogaSakai}.
When the square-triangle tilings are focused on,
the order metrics of triangular and square lattices are
smaller than the others, as summarized in Table~\ref{tbl}.
Here, we explore how the critical temperature
depends on the lattice structure as a secondary factor.

\begin{figure}[htb]
  \includegraphics[width=\linewidth]{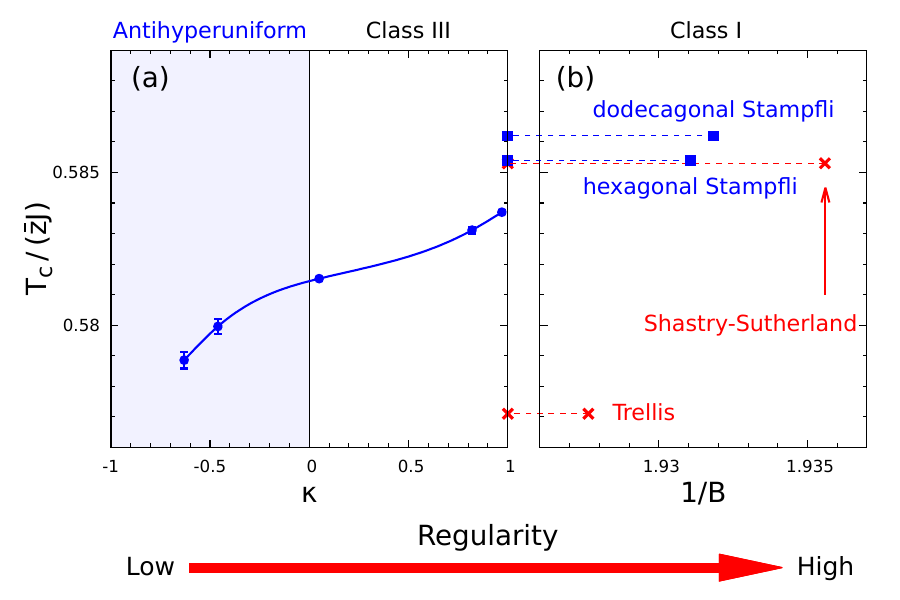}
   \caption{
     Critical temperatures of the square-triangle tilings
     as a function of (a) exponent $\kappa$ and (b) inverse of the order metric $1/B$.
     Blue circles and squares represent the results of the square-triangle and 
     Stampfli quasiperiodic tilings, respectively,
     all of which have $\bar{z}\simeq 5.07$.
     Red crosses represent the results of the trellis and Shastry-Sutherland lattices
     with $\bar{z}=5$.
   }
   \label{bTc}
 \end{figure}

Now, we focus on the systems with $\bar{z}\simeq 5.07$,
realized in the square-triangle and Stampfli quasiperiodic tilings.
The critical temperatures, shown as the blue symbols in Fig.~\ref{bTc},
demonstrate their dependence on the hyperuniform parameters.
As discussed earlier, in the square-triangle tilings,
increasing $\kappa$ causes a transition of the tiling patterns from antihyperuniform to hyperuniform,
which is accompanied by the increase of the transition temperatures,
as shown in Fig.~\ref{bTc}(a).
For Class I hyperuniform systems with $\kappa=1$ such as Stampfli quasiperiodic systems,
the order metric $B$ is useful to explore the relationship
between the regularity and critical temperature.
Figure~\ref{bTc}(b) shows that the critical temperature increases with $1/B$.
A similar trend is observed in the periodic systems with $\rho=5$,
which are shown by the red crosses,
supporting the significance of hyperuniformity.
Thus, the hyperuniform properties together with $\rho$ and $\bar{z}$
play a crucial role in determining the critical temperature of the Ising model.

Before summarizing the paper, we comment on the effect of bond dilutions
in the Ising model
since our square-triangle tiling can be regarded as 
the bond-diluted triangular lattice with the local rule~\cite{Kawamura_1983}.
According to Harris's criterion~\cite{Harris},
the random bond dilutions change the universality class of
the magnetic phase transition when $\alpha>0$,
where $\alpha$ denotes the exponent for the specific heat.
As a consequence, critical exponents are gradually altered by bond dilutions
in three dimensions~\cite{Hasenbusch_2007,Fytas_2010}.
However, in the two-dimensional case with $\alpha=0$,
the bond dilutions do not give any significant effect on the critical phenomena.
Although nontrivial logarithmic corrections have been proposed~\cite{Dotsenko,Shalaev,Shankar,Ludwig},
our analysis of the Ising model on square-triangle tilings did not detect such effects.
This may be attributable to an underlying order in the tiling pattern.

\section{Summary}\label{sec:summmary}

We have examined critical phenomena of the Ising models on the square-triangle tilings.
By using the growth rule with a parameter $p$,
we have systematically generated hyperuniform, nonhyperuniform,
and antihyperuniform square-triangle tilings.
Applying the classical MC simulations with Swendsen-Wang multi-cluster algorithm
to the Ising model on these tilings with distinct hyperuniform properties,
we have computed the magnetization, susceptibility, and Binder parameter.
Our results demonstrate that these quantities scale well with
the exponents of the two-dimensional Ising universality class
across different temperatures and radii.
Furthermore, we have compared the critical phenomena
in the Ising models on various tilings,
including square, triangular, Shastry-Sutherland, trellis, and Stampfli tilings,
all of which are composed of square and triangle tiles.
We have observed that, for a fixed average coordination number,
the critical temperature increases with
the degree of the regularity of the square-triangle tilings,
as quantified within the framework of hyperuniformity.
This finding suggests a promising approach
to controlling physical properties,
e.g. the critical temperature and coercivity in magnets,
by tuning the degree of regularity.

\begin{acknowledgments}
  Parts of the numerical calculations were performed
  in the supercomputing systems in ISSP, the University of Tokyo.
  This work was supported by Grant-in-Aid for Scientific Research from
  JSPS, KAKENHI Grant Nos. JP22K03525 (A.K.).
\end{acknowledgments}

\appendix

\section{Hexagonal and dodecagonal Stampfli tilings}\label{Stampfli}
\begin{figure}[htb]
  \includegraphics[width=\linewidth]{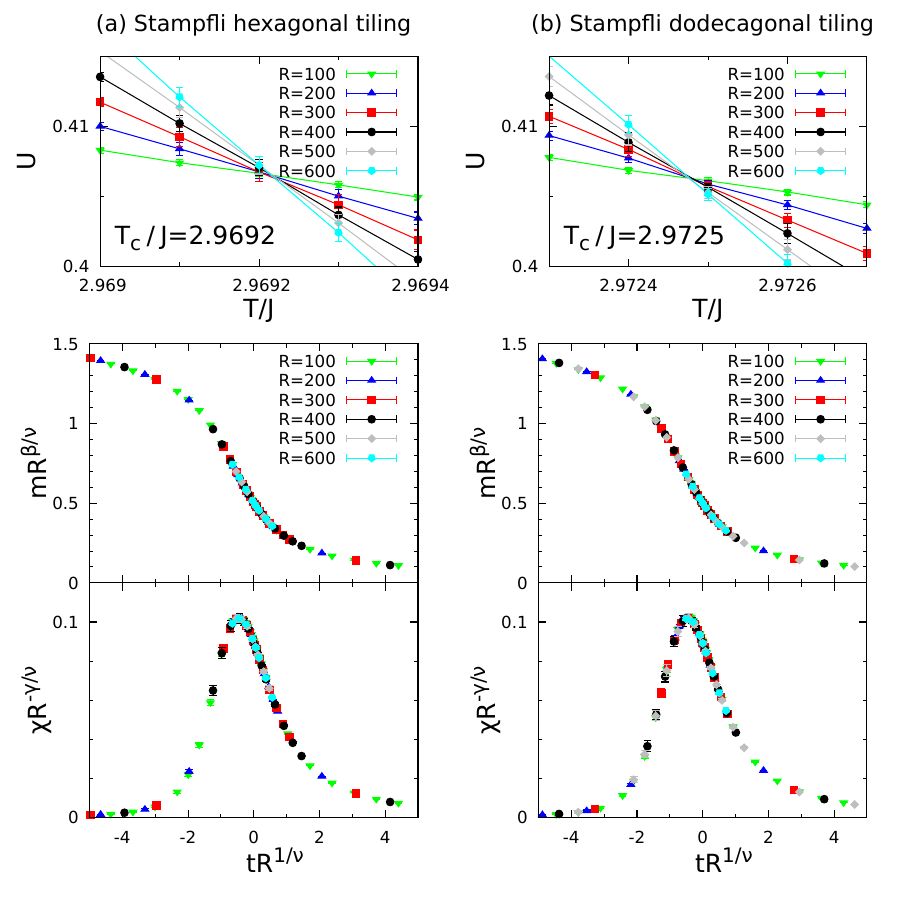}
  \caption{
    Binder parameter as a function of the temperature, and
    rescaled magnetizations and magnetic susceptibilities
    as a function of the rescaled temperature for
    the Ising models on (a) Stampfli hexagonal and (b) Stampfli dodecagonal tilings.
  }
  \label{stampfli}
\end{figure}

Here, we discuss critical behaviour in the Ising model 
on the Stampfli hexagonal and dodecagonal tilings.
We create the circular quasiperiodic tilings by means of the substitution rules.
Applying the classical MC simulations with SW algorithm 
to the Ising model on these tilings with distinct radii, 
we calculate Binder parameter, magnetization, and magnetic susceptibility.
The results are shown in Fig.~\ref{stampfli}.
We find that the crossing points of the lines of the Binder parameter
almost correspond to each other.
Then, we obtain the critical temperatures $T_c/J=2.9692$ and $2.9725$ 
for the Ising model on the Stampfli hexagonal and dodecagonal tilings, respectively.
We find that magnetizations and magnetic susceptibilities for distinct temperatures and radii are well scaled
by means of the exponents with $\beta=1/8, \gamma=7/4$, and $\nu=1$.
Therefore, critical phenomena of these models belong to the two-dimensional Ising universality class.

\nocite{apsrev42Control}
\bibliographystyle{apsrev4-2}
\bibliography{./refs}

\end{document}